\documentstyle[psfig,editedvolume]{crckapb}

\def\iras{IRAS\,13349+2438}  
\def\G{$\Gamma_{\rm x}$ }
\def\ros{{\sl ROSAT }}
\def\asca{{\sl ASCA }}

\def\approxlt{\mathrel{\hbox{\rlap{\lower.55ex \hbox {$\sim$}}
        \kern-.3em \raise.4ex \hbox{$<$}}}}
\def\approxgt{\mathrel{\hbox{\rlap{\lower.55ex \hbox {$\sim$}}
        \kern-.3em \raise.4ex \hbox{$>$}}}}

\begin{opening}
\title{Dusty warm absorbers: The case of IRAS\,13349+2438}  

\author{Stefanie Komossa}
\author{Dieter Breitschwerdt}
\institute{Max-Planck-Inst. extraterrestrische Physik, Postfach 1603, \\
           85740 Garching, Germany;~ skomossa@xray.mpe.mpg.de}
\author{Janek Meerschweinchen}
\institute{Weststrasse 19, 3063 Obernkirchen 2, Germany}

\end{opening}

\runningtitle{Dusty warm absorbers: The case of IRAS\,13349+2438}

\begin{document}


\vspace*{-10cm}
\begin{verbatim}
Contribution to the proceedings of `Astrophysical Dynamics'
(Evora, April 14-16, 1999); to appear in Ap&SS.
Preprint available at http://www.xray.mpe.mpg.de/~skomossa
\end{verbatim}
\vspace*{8.1cm}

\begin{abstract}
Warm absorbers are an important new probe of the
central region of active galaxies (AGN).
Observing and modeling this component provides a wealth of information on 
the nature of the warm absorber itself, its relation to other components
of the active nucleus, and the intrinsic AGN X-ray spectral shape.

We briefly review 
the general properties of dusty warm gas. 
For the first time, we then apply such a model to the IR loud quasar \iras. 
It was the first to be suggested to host a {\em dusty} warm
absorber (Brandt et al. 1996), but has not yet been modeled as such.    

\end{abstract}

\section{{\itshape Dusty} warm absorbers}

\subsection{ Influence of dust on the X-ray absorption structure}

So far, warm absorbers revealed their existence mainly in the soft X-ray spectral region
by imprinting absorption edges on the X-ray spectra of many
AGN. This highly ionized material provides an important new
diagnostic tool to investigate the physical conditions in the nuclei
of active galaxies.

Recently, evidence has accumulated that some warm absorbers
contain significant amounts of dust. This possibility was first
suggested by Brandt et al. (1996, B96 hereafter)
to explain the lack of excess X-ray {\em cold} absorption despite strong optical
reddening 
of the quasar IRAS 13349+2438.
A point emphasized by Komossa \& Fink (KoFi hereafter, e.g. 1997a,b) and
Komossa \& Bade (KoBa; 1998)
is
the strong influence of the presence of dust on the
X-ray absorption spectrum which becomes drastic for high column densities $N_{\rm w}$.
Signatures of the presence of (Galactic-ISM-like) dust, as opposed to
the dust-free case are, 
e.g., 
\begin{itemize}
\item  a strong carbon edge in the X-ray spectrum,
\item  a stronger temperature gradient across the absorber
with more gas in a `colder' state, shifting the dominant absorption
edges to more lowly ionized species
\item increased sensitivity of dusty gas
to radiation pressure, which may drive strong outflows of the warm material.
\end{itemize}
The first two effects lead to an {\em effective flattening} of the observed
X-ray spectrum. Thus, for both, the derivation of the physical
properties of the warm absorber as well as the intrinsic spectral shape
it is important to apply self-consistent models that take into account the
presence of dust.

\section{The IR-loud quasar IRAS\,13349+2438}

\begin{table}   
\begin{center}
  \caption{Properties of the dusty warm absorber in IRAS 13349 derived from X-ray spectral fits.
   For comparison results for a dust-free warm absorber are shown.
   The derived properties of the ionized material and the intrinsic continuum
   are rather different in this case. }
  \begin{tabular}{llcllll}
  \hline
     & \multicolumn{4}{l}{warm absorber} & \multicolumn{2}{l}{single powerlaw} \\
         & \G~~~ & log $U$~~ & log $N_{\rm w}$~~ & $\chi^2_{\rm red}$~~~~~ &
                                              \G~~~ & $\chi^2_{\rm red}$ \\
  \hline
    dusty WA  & --2.9 & --0.4 & 21.2$^{(1)}$ & 1.2 & --2.8 & 1.3 \\
    dust-free WA  & --2.2 &   ~0.7 & 22.7 & 0.8 & & \\
  \hline
     \end{tabular}
  \label{tab1}

  \noindent{ \footnotesize $^{(1)}$ fixed to
     the value $N_{\rm opt}$ determined from optical reddening }
\end{center}
\end{table}

This quasar received a lot of attention, recently.
A detailed optical study was presented by Wills et al. (1992).
In a thorough study of the \ros X-ray spectrum of \iras, 
B96 suggested the presence of a dusty warm absorber.
Brinkmann et al. (1996) detected changes in the \asca spectrum as compared to
the earlier \ros data;
the warm-absorption features remained present (Brandt et al. 1997).
Further candidates for dusty warm absorbers quickly followed,
including NGC\,3227 (KoFi 1996, 1997b), NGC\,3786 (KoFi 1997c), and IRAS\,17020+4544
(Leighly et al. 1997, KoBa). Whereas the model of a warm absorber that 
{\em includes} the presence of dust has been successfully fit to these
objects, it has not yet been applied to \iras~(first results of the
present study were reported by Komossa 1998, and Komossa \& Greiner 1999). 
Given the potentially strong modifications of the X-ray absorption spectrum
in the presence of dust,
it is important to  scrutinize whether a dusty warm absorber is consistent with
the observed X-ray spectrum.
Since some strong
features of dusty warm absorbers appear outside the \asca sensitivity range,
\ros data are best suited for this purpose; we used the two pointed PSPC observations
of \iras~of Jan. 1992 and Dec. 1992. 

As ionizing continuum illuminating the absorber we adopted a mean Seyfert
spectrum of piecewise powerlaws
(as in KoFi 1997b) with $\alpha_{\rm EUV}=-1.4$. We use as definition
for the ionization parameter $U = Q/(4\pi{r}^{2}n_{\rm H}c)$, where $Q$
is the rate of photons above the Lyman limit. The photoionization calculations
were carried out with Ferland's (1993) code {\em Cloudy}.

In a first step, we fit a dust-{\em free} warm absorber (as in B96,
but using the additional
information on the hard X-ray powerlaw available from the ASCA observation,
$\Gamma_{\rm x}^{\rm 2-10 keV} \simeq -2.2$).
This gives an excellent fit with log $N_{\rm w}$=22.7 ($\chi^2_{\rm red}$ = 0.8).
If this same model is re-calculated by fixing $N_{\rm w}$ and
the other best-fit parameters but {\em adding dust} to the warm absorber
the X-ray spectral shape is drastically altered and the data can not be
fit at all ($\chi^2_{\rm red}$ = 150). This still holds if we allow for
non-standard dust, i.e., selectively exclude either the graphite or silicate
species.

Whereas this first approach was mainly to demonstrate the strong influence 
of the presence of dust, it has to be kept in mind 
that the expected column density $N_{\rm opt}$ derived
from optical extinction is less than the X-ray value of $N_{\rm w}$ determined
under the above assumptions. Therefore, in a next step, we
allowed all parameters (except \G) to be free and checked, whether a dusty
warm absorber could be successfully fit at all.
This is not the case (e.g., if $N_{\rm w}$ is fixed to log $N_{\rm opt}$ = 21.2
we get $\chi^2_{\rm red}$ = 40).

The bad fit results can be partially traced back to the `flattening' effect of dust.
In fact, if we allow for a steeper intrinsic powerlaw spectrum, with \G $\simeq -2.9$
much steeper than the \asca value,
a dusty warm absorber with $N_{\rm w}=N_{\rm opt}$ fits the \ros
spectrum well ($\chi^2_{\rm red}$ = 1.2, Tab. 1).
We also analyzed the \ros all-sky survey data of this source and, again,
a steep intrinsic spectrum is required. 

\begin{figure}[t]
    \vbox{\psfig{figure=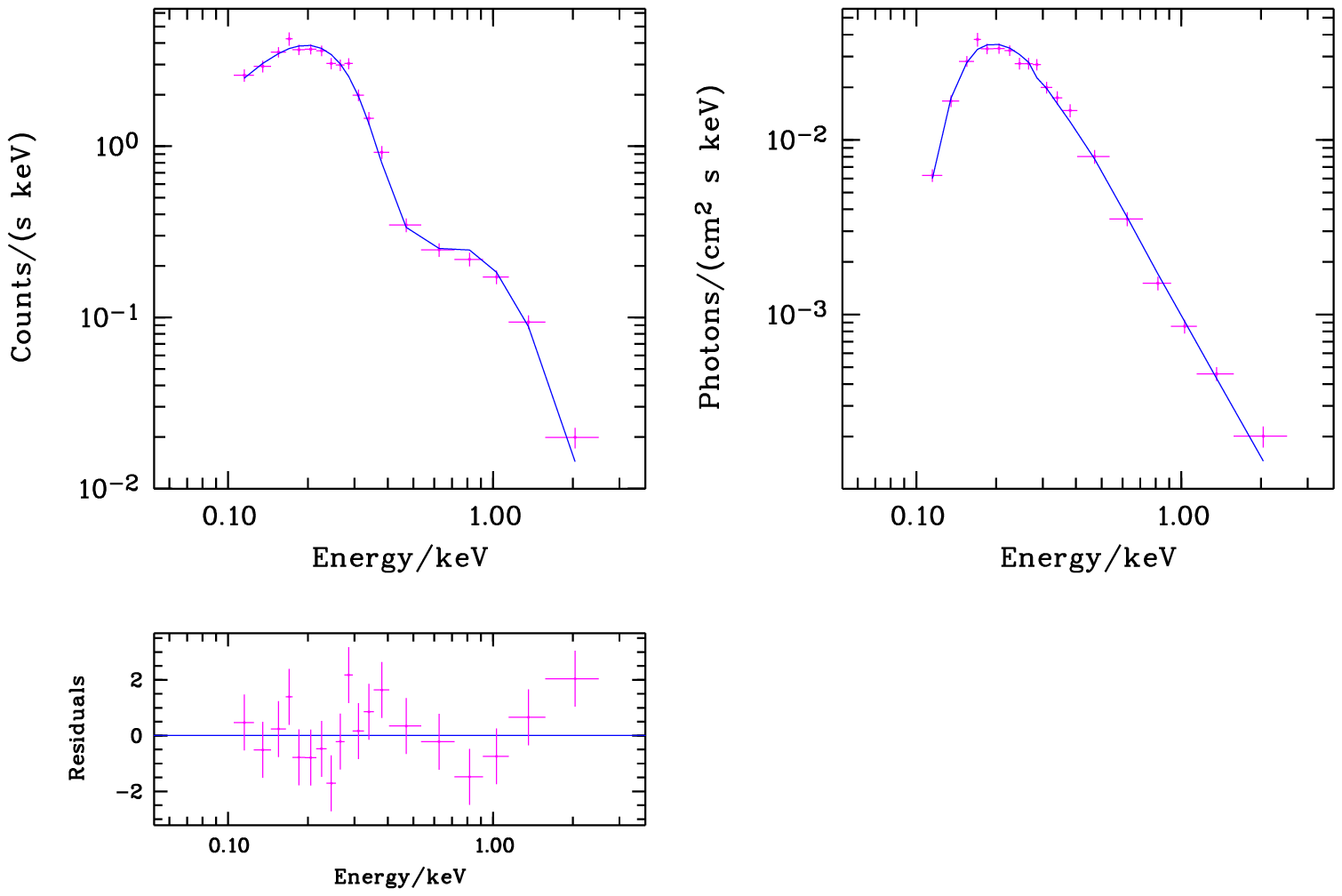,width=4.5cm,%
     bbllx=3.0cm,bblly=1.1cm,bburx=10.1cm,bbury=4.4cm,clip=}}\par
    \vspace*{-2.1cm}\hspace*{4.3cm}
    \vbox{\psfig{figure=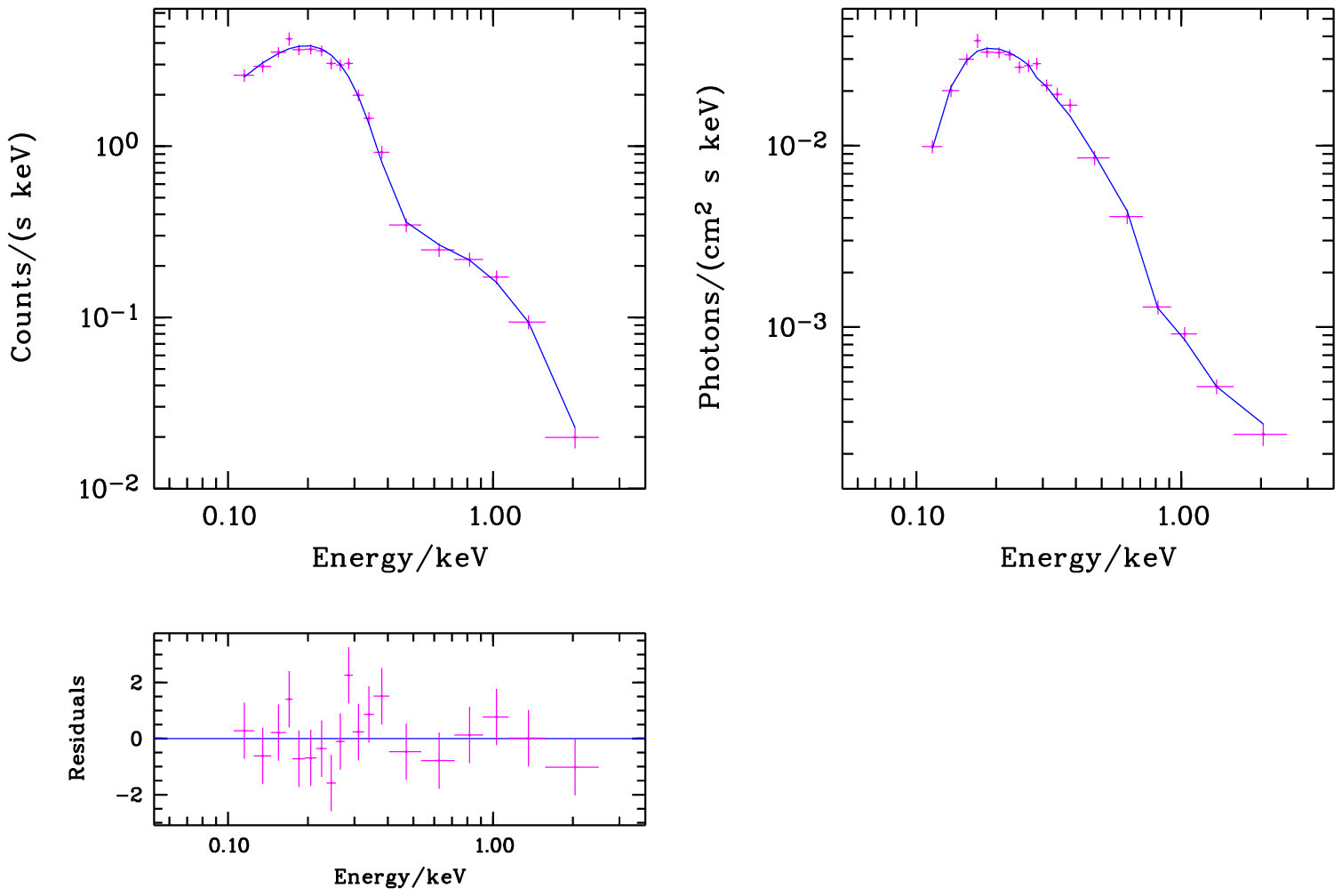,width=4.0cm,%
     bbllx=3.7cm,bblly=1.1cm,bburx=10.1cm,bbury=4.4cm,clip=}}\par
    \vspace*{-2.07cm}\hspace*{8.3cm}
    \vbox{\psfig{figure=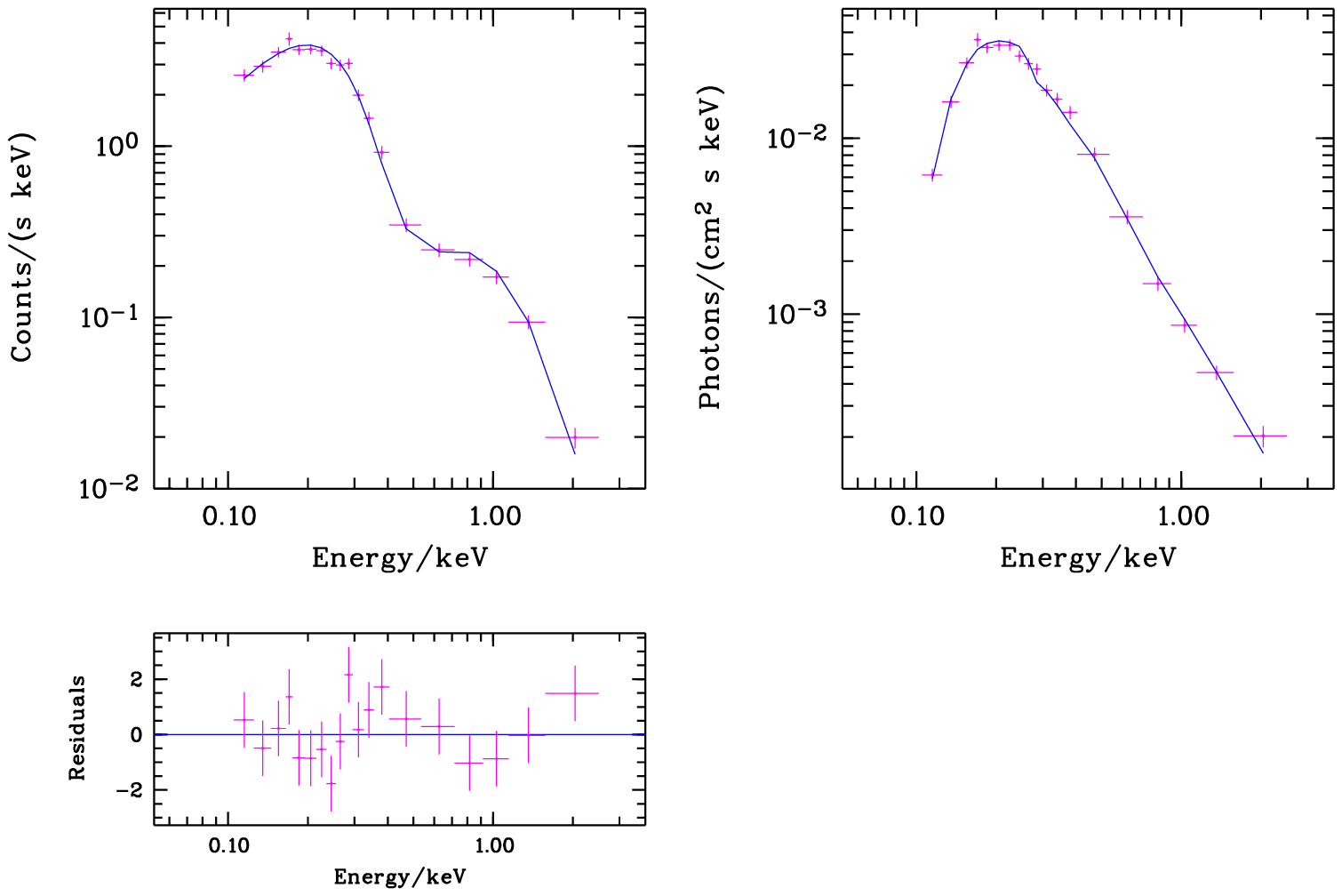,width=4.0cm,%
     bbllx=3.7cm,bblly=1.1cm,bburx=10.1cm,bbury=4.4cm,clip=}}\par
\vspace*{-0.4cm}
  \caption[]{Residuals after applying different spectral models to the
\ros X-ray spectrum of \iras. Left: simple powerlaw model, middle: dust-free warm
absorber, right: dusty warm absorber (see text for details). Systematic
residuals around 0.8 keV, the location of the OVII and OVIII absorption edges,
are better removed by the dust-free warm absorber due to the presence of
a much stronger OVIII edge in the best-fit model as compared to the dusty
case. Fine-tuning in the dust-properties or the oxygen metal abundances
with corresponding changes in the heating-cooling balance might alleviate
this problem. }
\end{figure}

At present,
there are several possible explanations for the {\sl ROSAT}-\asca spectral differences:
(i) Variability in the intrinsic powerlaw. We consider this solution unlikely,
since the spectrum is steep in all three \ros observations taken at different times.
(ii) Remaining {\sl ROSAT}-\asca cross-calibration uncertainties. 
This is well possible, given
previous reports of a tendency of {\sl steeper} \ros spectra when compared
with \asca data of the same object.   
(iii) Variability in a {\em two}-component warm absorber. There are increasing
indications of more than one X-ray warm absorber in MCG\,6-30-15, and
\iras~might be a similar case.   

In summary, the presence of a {\em dusty} warm absorber 
in IRAS\,13349+2438 seems consistent with the data if one allows for some spectral differences
between the \ros and \asca data. 
The necessity of dust survival then locates the ionized material
{\em outside} the bulk of the broad line region (BLR) of \iras.

\section{Prospects for the study of dusty WAs with {\sl XMM} and {\sl AXAF}}

Clear signatures of the presence of dust are the carbon edge at 0.28 keV
and the oxygen edge at 0.56 keV (not yet individually resolved by current X-ray
instruments), and the confirmation and study of dusty warm absorbers
will certainly
be a prime goal of future X-ray satellites like {\sl AXAF} 
and {\sl XMM}.
In particular,
{\em dust}-created absorption edges will play an important role not
only in probing components of the active nucleus, like the dusty torus,
but also are they a very useful diagnostic
of the dust properties in other galaxies{\footnote{Note
that, if the dust was mixed with {\sl cold} gas instead, the soft X-ray spectrum
would be completely dominated by hydrogen absorption (in {\sl warm} absorbers, H and He are
nearly fully ionized). Further, gas-phase absorption would be very difficult
to distinguish from dust-absorption, since the differences in
edge energies of {\sl neutral} atoms due to solid-state effects in
the dust are only on the order of a few eV (cf. the discussion in KoFi 1997a and KoBa).}}.
Further, measurements of edge energies and widths will allow to determine the
dynamical state of the warm absorber.

{}

\end{document}